# Miniaturized Double-Wing Delta-E Effect Sensors

*Fatih Ilgaz, Elizaveta Spetzler, Patrick Wiegand, Franz Faupel, Robert Rieger, Jeffrey McCord, Benjamin Spetzler\**

F. Ilgaz, Franz Faupel

Multicomponent Materials, Department of Materials Science, Faculty of Engineering, Kiel University, 24143 Kiel, Germany

E. Spetzler, J. McCord

Nanoscale Magnetic Materials - Magnetic Domains, Department of Materials Science, Faculty of Engineering, Kiel University, 24143 Kiel, Germany

P. Wiegand, R. Rieger

Networked Electronic Systems, Department of Electrical and Information Engineering, Faculty of Engineering, Kiel University, 24143 Kiel, Germany

B. Spetzler

Micro- and Nanoelectronic Systems, Department of Electrical Engineering and Information Technology, Ilmenau University of Technology, 98693 Ilmenau, Germany
E-mail: benjamin.spetzler@tu-ilmenau.de

Magnetoelastic composites are integral elements of sensors and actuators utilizing magnetostriction for their functionality. Their sensitivity typically scales with the saturation magnetostriction and inversely with magnetic anisotropy. However, this makes the devices prone to minuscule residual anisotropic stress from the fabrication process, impairing their performance and reproducibility, hence limiting their suitability for arrays. This study presents a shadow mask deposition technology combined with a free-free magnetoelectric microresonator design intended to minimize residual stress and inhomogeneity in the magnetoelastic layer. Resonators are experimentally and theoretically analyzed regarding local stress anisotropy, magnetic anisotropy, and the ΔE effect in several resonance modes. Further, the sensitivity is analyzed in the example of ΔE-effect sensors. The results demonstrate a device-to-device variation of the resonance frequency < 0.2 % with sensitivities comparable with macroscopic ΔE-effect sensors. The reproducibility is drastically improved over previous magnetoelastic device arrays. This development marks a step forward in the reproducibility and homogeneity of magnetoelastic resonators and contributes to the feasibility of large-scale, integrated sensor arrays.

## 1. Introduction

In today's technologically advanced world, magnetic field sensors and actuators[1] have become essential components across a diverse range of industries, including aerospace,[2] automotive,[3] electronics,[4] and biomedical applications.[5–7] In recent years, thin-film magnetoelectric (ME) sensors have become a class of promising magnetometers for detecting low-frequency and small-amplitude magnetic fields.[8–13] These sensors comprise magnetoelectric composites consisting of mechanically coupled magnetostrictive and piezoelectric components. They can be easily integrated with electronics and have the potential for array-based applications.[14,15] Magnetoelectric sensors have demonstrated detection limits in the low pT regime through the direct magnetoelectric effect; however, their operation is either limited to macroscopic sizes of the sensor elements[16] or to high frequencies and narrow bandwidths of a few Hz around the sensor's resonance frequency.[17] These limitations can be overcome by utilizing a modulation technique based on the ΔE effect, which is the change in the mechanical stiffness tensor of a magnetostrictive material upon applying a magnetic field due to additional magnetostrictive strain.[18,19] The change in the stiffness tensor induces a shift in the sensor's resonance frequency, which can be read out electrically.[20] ΔE-effect sensors designed as plate and cantilever resonators have demonstrated detection limits in the sub-nT regime at frequencies ranging from 10 Hz to 100 Hz.[12,21–25]

Previously investigated cantilever-type mm-sized ΔE-effect sensors have encountered significant challenges related to inhomogeneous magnetic properties at the clamping region caused by residual stress and shape anisotropy.[26–28] Moreover, the larger dimensions of the magnetoelastic resonators limit spatial resolution and make them unsuitable for compact array configurations.[29] These limitations were attempted to be solved by μm-sized counter-mode resonators.[30] However, a persistent challenge that arises in such sensors is their reproducibility. These sensors exhibit a huge device-to-device performance variation of more than a factor of two difference in the frequency response between sensors with exact dimensions caused by minimal stress that couples into the magnetic properties via the large magnetostriction.[30] Using large effective anisotropies or small magnetostriction decreases the influence of stress on magnetic properties but simultaneously reduces the sensor's sensitivity.[26] This variation in performance affects sensor yield and leads to undefined sensor properties, making it difficult to achieve consistent and reliable sensor performance. Developing a reliable technology for fabricating magnetoelastic sensors is crucial to address these issues. Typically, the samples are produced by depositing and structuring the magnetoelastic layer on a constrained resonator, which is subsequently etched out (released).[21] During this process, stress is unintentionally introduced into the magnetoelastic layer via the relaxation of intrinsic stress in the substrate and other layers.[30] Solving the reproducibility problem is of utmost importance for large-scale industrial applications and the cost-effective fabrication of magnetoelastic devices. By developing a reliable fabrication technology and minimizing the variability in device performance, the full potential of magnetoelastic resonators can be realized, enabling their widespread adoption in various industries and research fields.

In this work, we present a shadow-mask deposition technology and demonstrate the extremely high reproducibility of magnetoelectric thin-film resonators. Further, we develop a double-wing micro-resonator design with various advantages compared to classical cantilevers. The results represent a significant step toward reliable fabrication of magnetoelastic resonators, cheap, large-scale fabrication, and device arrays. First, we demonstrate the micro-resonator design and fabrication technique. Then, we show the induced spatial stress distribution in a representative resonator upon the magnetic layer deposition, combining optical measurements and a magneto-mechanical model. Further, we employ magneto-optical Kerr effect (MOKE) microscopy to study the magnetic properties of the sensor. We utilize finite element method (FEM) simulations to identify different resonance modes and validate the corresponding resonance frequencies and mode shapes with vibrometer measurements. We conduct admittance measurements to determine the resonance frequency shift and relative magnetic and electrical sensitivities for the identified resonance modes. Finally, we explore arrays comprising many parallel-connected magnetoelastic sensor elements with identical geometries to investigate the reproducibility and compare their performances with single sensors.

## 2. Results and Discussion

### 2.1 Resonator Design and Technology

The individual resonators (**Figure 1**a,b) are based on electromechanical thin-film multilayer structures comprising a 10-µm-thick doped poly-Si substrate, also functioning as a rear-side electrode, with a 0.2-µm-thick pad oxide layer on the top to insulate it electrically from the subsequent layers. On top of the pad oxide layer, a 0.5-µm-thick AlN piezoelectric layer is deposited, followed by two 1-µm-thick patterned Al electrodes symmetrically placed on both sides of the anchors for actuation and read-out. A top view of an example resonator is shown in Figure 1d. Finally, a 200-nm-thick amorphous magnetostrictive layer $(Fe_{90}Co_{10})_{78}Si_{12}B_{10}$ (FeCoSiB) is deposited on the rear side of the free-standing resonators.

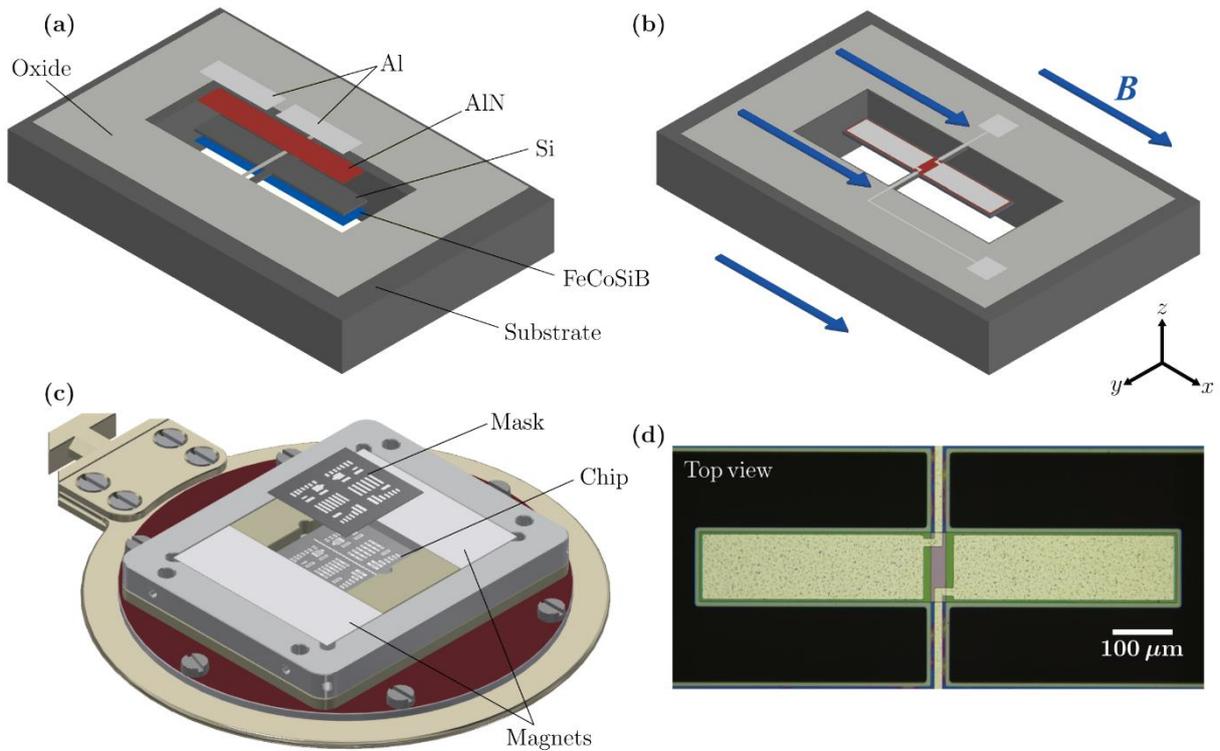

**Figure 1.** (a) Schematic of a miniaturized double-wing ΔE-effect sensor with the individual layers. (b) Schematic view of the complete sensor during operation with the indicated direction of the applied magnetic field during measurements. (c) Illustration of the magnetic layer deposition with shadow-mask deposition technique. (d) Optical microscopy image of the top of an example resonator.

In contrast to previously used lithography methods,[21,30–32] the magnetostrictive layer is deposited through a shadow mask onto the rear side of the released resonators using an in-house built magnetron deposition system (sample holder in Figure 1c). The setup permits micrometer-precision mask alignment for shadow masks with sub-µm feature size. No additional lithography processes for structuring the magnetic layer were used. Because the magnetic layer is structured via the shadow mask and deposited as the last layer on the released resonator, the residual stress in the magnetic layer is only determined by the magnetic layer deposition process and not by the fabrication of the other underlying layers or the substrate. This reduces the degrees of freedom to be controlled during the fabrication and permits precise stress control by adjusting the deposition conditions (see Experimental Section). During the deposition of the magnetic layer, a magnetic field is applied by permanent magnets to induce a magnetic easy axis along the short axis of the resonator.

A second component contributing to the reproducibility and performance is the resonator design. Notable technological advantages result from using double-wing micro-resonators (Figure 1b) instead of a classical cantilever geometry. Anchoring the double-wing resonator in the center permits homogeneous magnetic layer deposition on the entire resonator. It avoids partial shadowing of the geometry by the substrate, which would occur for cantilevers at the clamping. As we will show in Section 2.4, an antisymmetric resonance mode can be excited with the proposed resonator design, which reduces the detrimental influence of magnetic inhomogeneities at the anchor region and the tips. Additionally, the anchor design reduces clamping loss compared to a cantilever design[26,27] and, thereby, the coupling of adjacent resonator elements via the substrate.

For this work, we fabricated magnetoelastic resonators with various lengths of 400-850 µm and widths of 60-125 µm. The microfabrication and deposition processes are detailed in the Experimental Section. All results presented in the following section are for a representative resonator ID1 with in-plane dimensions of 640 µm × 105 µm. The data for all other produced samples is available in Supporting Information.

## 2.2. Residual Stress

To demonstrate the stress control of our deposition technology, we analyze the residual stress induced during the deposition of the magnetic layer and its influence on the effective magnetic anisotropy. For that, the out-of-plane displacement $u_z$ of the representative resonator with in-plane dimensions of 640 µm × 105 µm was measured with a laser profilometer before and after the FeCoSiB deposition. A mechanical finite-element-method (FEM) model is fitted to the measured data to identify intrinsic stress in the substrate and the magnetic layer. **Figure 2** shows the measured and simulated $u_z$ before (Figure 2a-c) and after (Figure 2d-f) the FeCoSiB deposition. The resonator is slightly bent upwards before the FeCoSiB deposition owing to the relaxation of residual stress in the nonmagnetic layers upon release of the resonator. Simulations and measurements match very well, assuming an isotropic initial stress in the substrate of $-140$ MPa (Figure 2a-c).

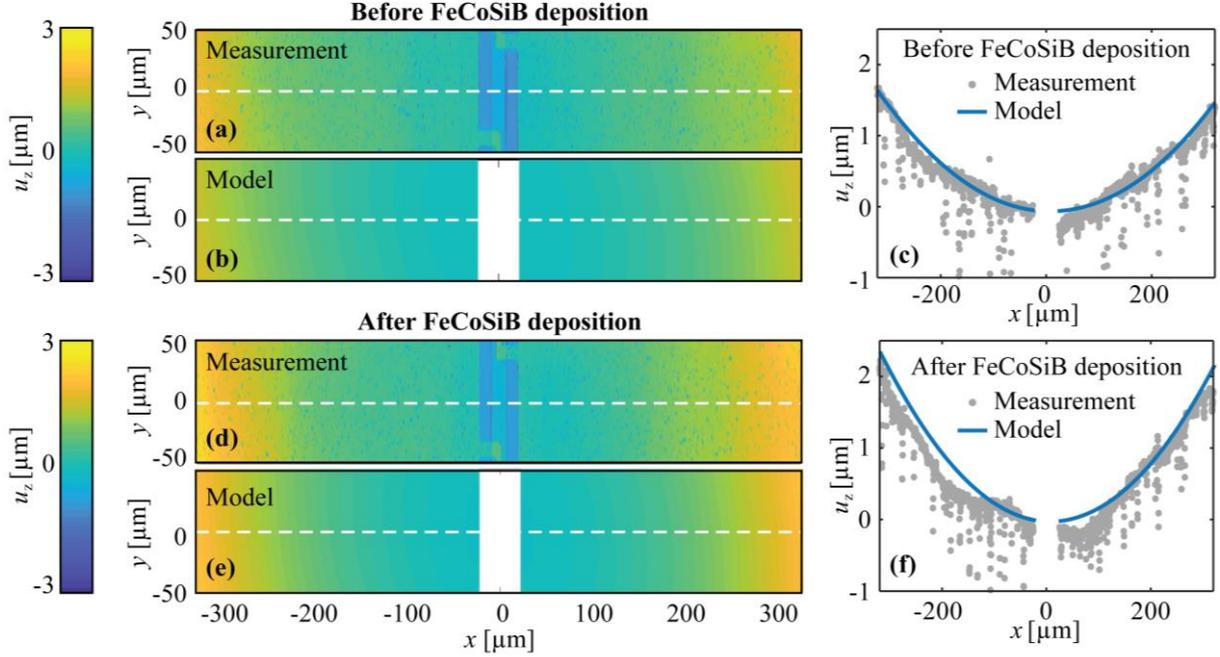

**Figure 2.** Measured and simulated z-displacement $u_z$ of a representative sensor (a-c) before and (d-f) after FeCoSiB deposition. (c) and (f) show the data from the cut lines marked with white dashed lines in figures (a-b) and (d-e), respectively. For the displacement simulation before FeCoSiB deposition (a-c), an initial isotropic stress of -140 MPa was applied to the substrate, and after the deposition, an additional homogeneous initial stress of $\sigma_{11} = -245$ MPa and $\sigma_{22} = -235$ MPa was applied to the magnetic layer.

After the deposition of the magnetic layer, the displacement of the resonator wings is slightly increased by approximately 0.5 µm at the tips (Figure **2**c,f), indicating deposition-induced compressive stress in the magnetic layer. The simulations and measurements match very well (Figure 2d-f) for a homogeneous initial stress of $\sigma_{11} = -245$ MPa and $\sigma_{22} = -235$ MPa with a minuscule anisotropy of $\sigma_{11} - \sigma_{22} \approx -10$ MPa. Because of the shape of the cantilever, the stress anisotropy $\sigma_{11} - \sigma_{22}$ decreases slightly after the relaxation, reaching an equilibrium value of $\sigma_{11} - \sigma_{22} \approx 9.3$ MPa in the center of the wings (see Supporting Information). As a result, the stress-induced magnetic anisotropy in the magnetic layer reaches $K_\sigma \approx 450$ Jm$^{-3}$ (saturation magnetostriction constant $\lambda_s = 30$ ppm[33]) with the magnetic easy axis oriented along the short axis (y-axis) of the resonator.

To validate the estimation of the stress anisotropies, we employed magneto-optical Kerr effect (MOKE) microscopy.[34] **Figure 3**a shows the domain configuration of the sample after demagnetization along its long axis. The magnetic domains orient predominantly along the short axis of the cantilever. They bend slightly around the clamping region because of the shape anisotropy and different stress relaxation at the edges compared to the center of the resonator (see Figure S1 in Supporting Information).

We extracted local magnetization curves from additional MOKE measurements to estimate the distribution of the local differential magnetic susceptibility $\chi$ at zero field along the long axis (x-axis) of the sample. The results are shown in Figure 3b. In the center of the resonator, the differential susceptibility is $\chi \approx 950$; it decreases close to the edges to values of $\chi < 400$. To quantify the individual energy contributions that define the value of $\chi$, we use a macrospin model. In the model we consider magnetoelastic anisotropy energy, demagnetizing field energy, and uniaxial magnetization-induced anisotropy energy introduced during sputtering along the short cantilever axis. FEM simulations are performed to obtain the demagnetizing field energy, while magnetoelastic anisotropy energy density $K_\sigma$ is taken from the residual stress analysis. The magnetization-induced anisotropy energy density $K_M$ is considered a fitting parameter. The simulations of $\chi$ match the measurements (Figure 3b) for a realistic (and spatially constant) value of $K_M \approx 500$ Jm$^{-3}$.[35] The simulations also confirm that the decrease of $\chi$ at the edges is caused by the demagnetizing field (see Supporting Information for details). Hence, the residual stress analysis is overall consistent with the measured magnetic properties.

Local magnetization curves were recorded on the beam and on the anchors to distinguish their local magnetic behavior (Figure 3c). The magnetization curve recorded on the anchors exhibits a significantly different shape, indicating a distribution of the effective magnetic anisotropy compared to the beam region. The reduced slope at $B > 0.5$ mT is mainly caused by the demagnetizing field (see Figure S2a, Supporting Information). Similar magnetic properties have been observed for all other resonators produced for this paper, as shown in the Supporting Information. The high inhomogeneity and effective anisotropy of the magnetic properties in the anchor region are expected to deteriorate the sensor performance if this region is active during sensor operation.[22,24,26,27] In the next section, we will show how a careful selection of the resonance mode based on the known distribution of the magnetic properties can improve the frequency tunability.

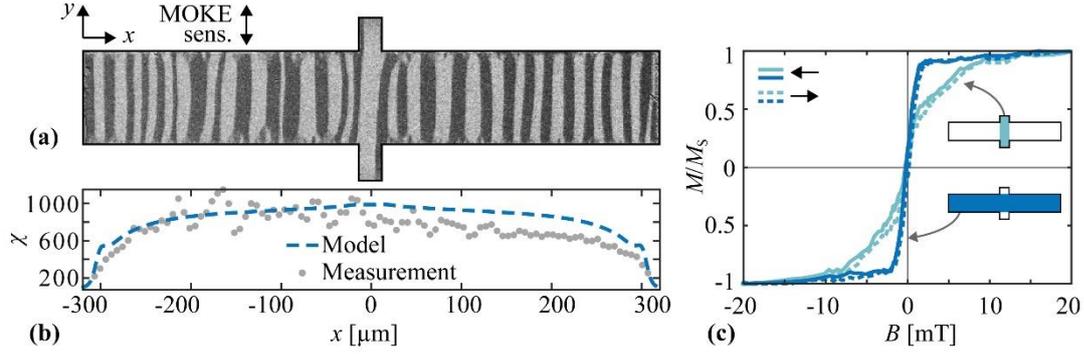

**Figure 3.** Magnetic properties of the resonator ID1 obtained by magneto-optical Kerr effect (MOKE) microscopy. (a) MOKE image of the magnetic domains of the sensor after demagnetizing along the x-axis. The magneto-optical sensitivity axis (MOKE sens.) is aligned with the y-axis as indicated. (b) Spatial distribution of the differential magnetic susceptibility $\chi$ estimated from the measurements and compared with simulations. (c) Local magnetization curves measured in two different regions with $B$ applied along the long axis of the resonator.

## 2.3. Resonance Modes

To analyze the performance of the exemplary resonator as a ΔE-effect sensor, we selected the first four resonance modes (RM1-4) using FEM simulations and vibrometer measurements. The measured and simulated mode shapes (at $B = 0\,\text{mT}$) are shown in **Figure 4**a-c, with eigenfrequencies of approximately 125.1 kHz (RM1), 366 kHz (RM2), 685.4 kHz (RM3) and 1.3 MHz (RM4). Corresponding frequencies of these modes for other produced resonators are available in Table S1 in Supporting Information. The simulations match the measurements very well, with minor deviations of the resonance frequencies $f_r$ smaller than 1.3 %, except for RM2, with a deviation of 5.5 %. All four modes are of a first or higher order bending type and differ in their displacement profiles and dynamic stress distributions (Figure 4d). Among the four resonance modes, only RM3 is asymmetric, with a minimum magnitude of $\sigma_{11}$ between the two anchors and a maximum $|\sigma_{11}|$ in the center of the resonator wings. In all other modes, it is $|\sigma_{11}| > 0$ in the anchor region. The different spatial distribution of the dynamic stress in the resonance modes will allow us to minimize the influence of undesired local magnetic properties by selecting a suitable mode in the next subsection.

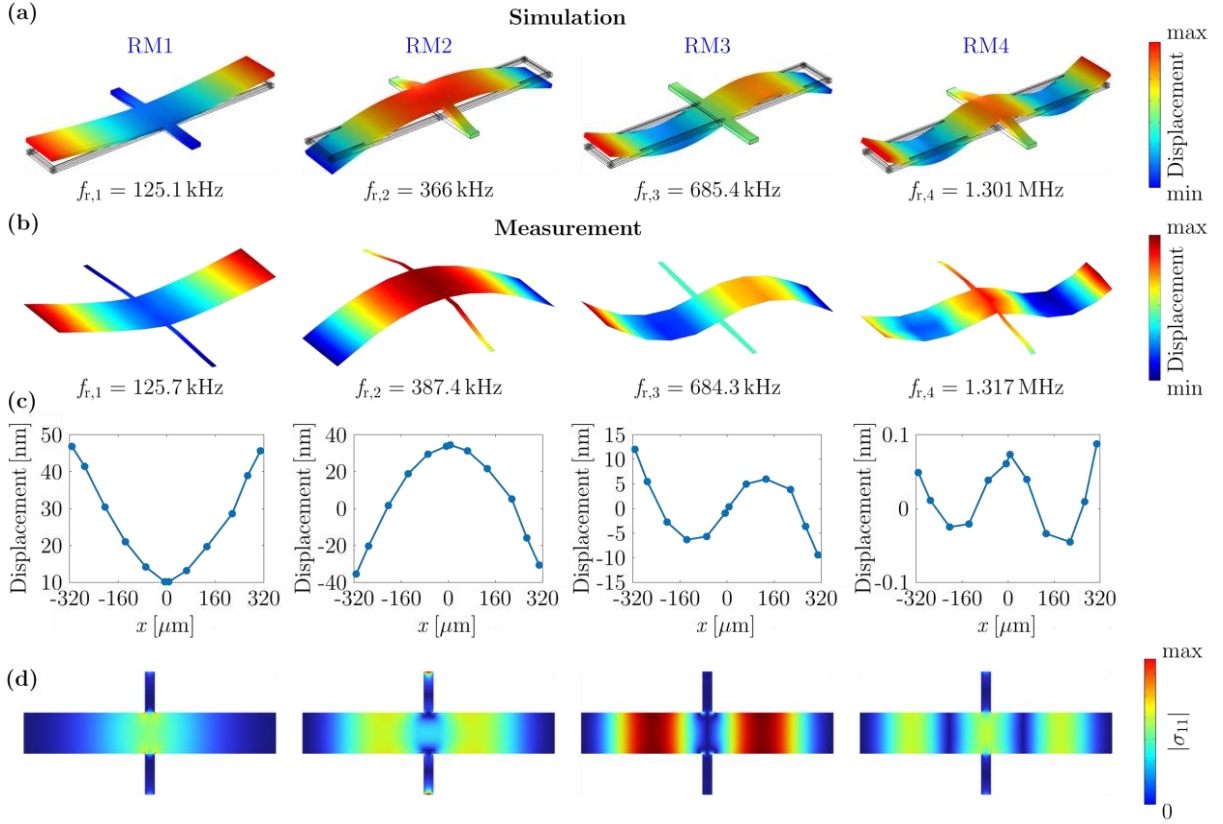

**Figure 4.** (a) Mode shapes of the first four resonance modes and their corresponding resonance frequencies at the example of sensor ID1 simulated with a FEM model, and (b) determined by vibrometer measurements. (c) Measured out-of-plane displacements along the center of the long axis of the sensor. (d) Distributions of the simulated $\sigma_{11}$ component of the stress tensor at the center of the magnetic layer.

## 2.4. Sensitivity and ΔE Effect

One of the main characteristics of a magnetic field sensor is its sensitivity to magnetic fields. As a measure for the sensitivity of the resonator as a ΔE-effect sensor, we use the amplitude sensitivity $S_{am} := S_{m,r} \cdot S_{el,r}$, as defined previously.[27] It is proportional to the change in resonance frequency $f_r$ induced by the applied magnetic flux density $B$ via the ΔE effect and to the slope of the sensor admittance $Y(f)$. These two proportionality factors are normalized to the operating frequency and referred to as relative magnetic sensitivity $S_{m,r}$ and relative electric sensitivity $S_{el,r}$, respectively. Details are provided in the Experimental Section. In the following, we first analyze the magnetic and electric sensitivities of the different resonance modes and then combine them to draw conclusions about the amplitude sensitivity.

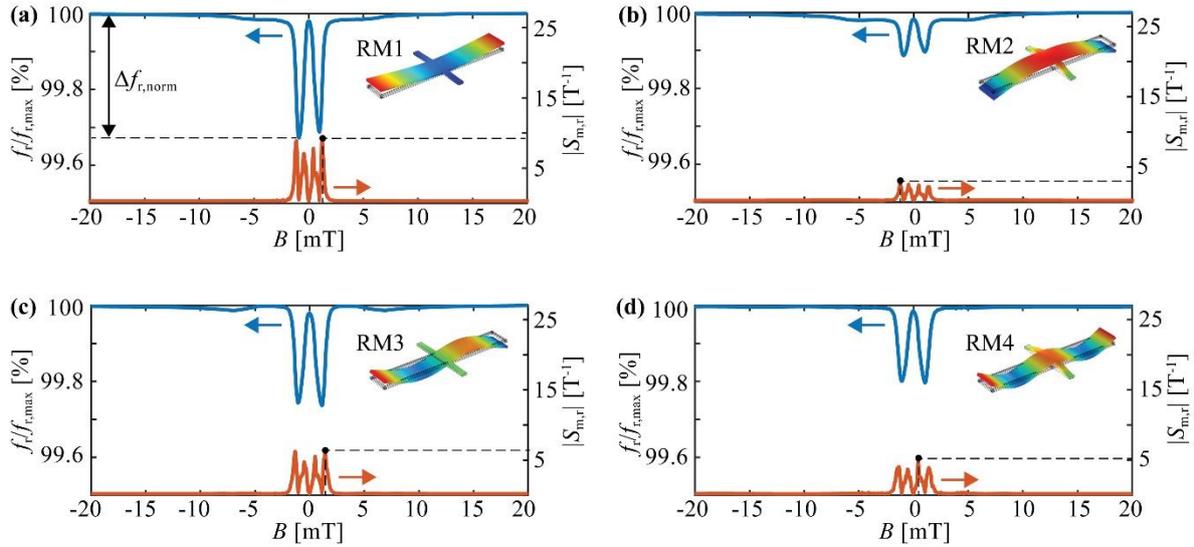

**Figure 5.** Measured normalized resonance frequencies $f_r/f_{r,max}$ and relative magnetic sensitivities $S_{m,r}$ as functions of the magnetic flux density $B$ applied along the long axis of the resonator for (a) RM1 with $f_{r,max} = 125.7$ kHz, (b) RM2 with $f_{r,max} = 366$ kHz, (c) RM3 with $f_{r,max} = 685.4$ kHz, and (d) RM4 with $f_{r,max} = 1302$ kHz. Magnetic working points are indicated with black dots. The arrows indicate the respective y-axis of the plotted data sets. The normalized frequency detuning $\Delta f_{r,norm}$ is indicated with an arrow in (a) as an example.

### 2.4.1. Magnetic Sensitivity and ΔE Effect

The dependency of the normalized resonance frequency $f_r(B)/f_{r,max}$, on the applied magnetic flux density $B$ is shown in **Figure 5** for RM1-4. All four curves are overall w-shaped, typical for bending mode resonators with an effective magnetic anisotropy perpendicular to the main dynamic stress axis and the applied magnetic field.[21,26,27] However, quantitative differences are apparent.

The normalized frequency detuning $\Delta f_{r,norm} := (f_{r,max} - f_{r,min})/f_{r,max}$, i.e., the difference between the maximum resonance frequency $f_{r,max}$ and the minimum resonance frequency $f_{r,min}$ differs significantly between the four resonance modes. It is largest in RM1 with $\Delta f_{r,norm}$ of 0.31 %, followed by RM3 with 0.26 % and RM4 with 0.2 %. The smallest change of 0.11% is measured in RM2. This leads to correspondingly different relative magnetic sensitivities $S_{H,r}$, as shown in Figure 5, with indicated maximum values $S_{m,r} = 8.9$ T$^{-1}$ (RM1), $S_{m,r} = 2.7$ T$^{-1}$ (RM2), $S_{m,r} = 6.2$ T$^{-1}$ (RM3), and $S_{m,r} = 4.8$ T$^{-1}$ (RM4). The same trend is visible for other produced resonators (see Table S2 in Supporting Information).

The resonance modes also differ in their saturation behavior. RM1, RM2, and RM4 show a minuscule increase in the resonance frequency at $|B| > 5$ mT before they reach their maximum values, while the resonance frequency in RM3 slightly drops to a local minimum at ≈ 7 mT before increasing again at larger flux densities.

All these differences in the resonance frequency curves can be well explained by the spatially varying effective magnetic anisotropy. The effective magnetic anisotropy is locally weighted by the alternating stress field of the respective resonance mode, and thereby, leads to resonance-mode dependent resonance frequency curves.[27]

The resonance frequency of RM2 (Figure 5b) is dominated by the elastic properties of the anchors, where the effective magnetic anisotropy of the magnetic layer is comparatively large and highly inhomogeneous (Section 2.2). Additionally, the magnetic layer thickness on the anchors is expected to decrease radially because the anchors are shadowed by the frame during the magnetic layer deposition. Both factors cause the small frequency detuning of RM2. Consistently, the best performance is obtained in RM1 and RM3, where the contributions of the anchors to the resonance frequency is minor. In RM3, the resonance frequency detuning is slightly smaller than in RM1. This is likely caused by a small contribution of the $C_{66}$ stiffness tensor component via the shear stress component $\sigma_{12}$ at the anchors (see Supporting Information), which are twisted during the oscillation in RM3. As demonstrated previously, the superposition of the ΔE effect in the shear component ($C_{66}$) and the longitudinal component ($C_{11}$) of the stiffness tensor compensate slightly and can result in an overall reduced frequency detuning compared to a pure bending mode.[27] The contribution of this shear-stress component to the resonance frequency of RM3 explains the different saturation behavior of RM3 compared to the other resonance modes. The slightly increasing resonance frequency at $|B| > 5$ mT visible in RM1, RM2, and RM4 is caused by the residual nonzero susceptibility visible in the magnetization measurements (Section 2.2). In RM3, the increase in resonance frequency is not visible because it is superposed by the contribution of the shear component, which causes the local minimum in the resonance frequency curve at ≈ 7 mT.

*2.4.2. Electric sensitivity and amplitude sensitivity*

Admittance characteristics of RM1-RM4 are shown in **Figure 6**a-d as functions of the normalized frequency $\Delta f/f_\mathrm{r}$ with $\Delta f = f - f_\mathrm{r}$ at their respective magnetic working point. The admittance magnitude $|Y|$ is normalized to its value $|Y_0|$ at the resonance frequency $f_\mathrm{r}$ for easier comparison of the data. The relative electric sensitivity is obtained from the derivative of the admittance and plotted in Figure 6a-d as well. The largest relative electric sensitivity is reached for RM3 with $S_{\mathrm{el,r}} = 19.5$ mS. This is a factor of approximately 25 times the maximum electrical sensitivities of RM1, RM2, and RM4 with values of $S_{\mathrm{el,r}} = 0.8$ mS, $S_{\mathrm{el,r}} = 4$ mS, and $S_{el,r}$ = 2.8 mS. A similar trend was observed for most of the other produced resonators (see Table S2 in Supporting Information). This is expected because the electrodes are specifically optimized for RM3, i.e., they cover the centers of the two wings but only partially cover the center of the resonator between the anchors where the other resonance modes are most active. As a result, the largest amplitude sensitivity is reached in RM3 with $S_{am}$ = 121 µS mT$^{-1}$. This sensitivity is similar to those previously reported for mm-sized ΔE-effect sensors.[21,23,24,26,27,29,36]

An mBvD model is fitted to the measurements to extract the quality factors of the resonance modes. With values of 600 (RM1), 662 (RM2), 765 (RM3), and 394 (RM4), they are slightly smaller than in previously investigated mm-sized ΔE-effect sensors.[24,29,37] As a consequence, the bandwidth $f_{BW} = f_r/Q$ of the microresonator sensors is up to two orders of magnitudes larger due to orders of magnitude higher resonance frequencies. The bandwidths for RM1, RM2, and RM3 are $f_{BW}$ = 200 Hz, $f_{BW}$ = 532 Hz, and $f_{BW}$ = 842 Hz, respectively. RM4 shows the largest bandwidth of $f_{BW}$ = 3.4 kHz due to its highest $f_r$ and smallest $Q$ value.

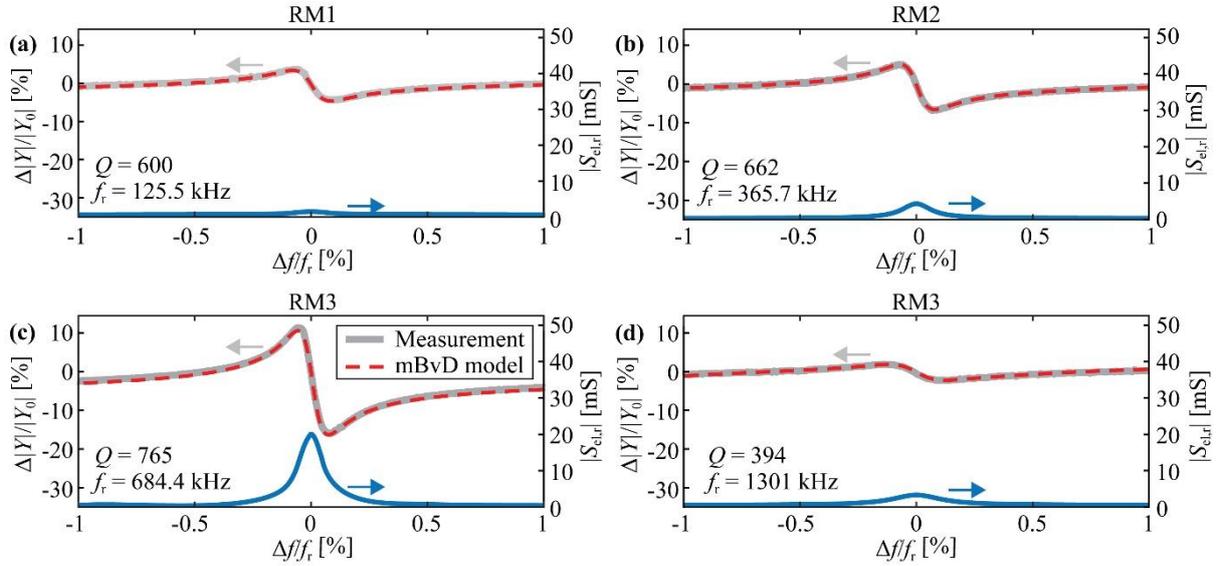

**Figure 6.** Measured and modeled normalized admittance magnitude $|Y|/|Y_0|$ and relative electrical sensitivity $S_{el,r}$ at the magnetic working point as a function of normalized excitation frequency $\Delta f/f_r$ (with $\Delta f = f - f_r$) for the first four resonance modes (a) RM1 with $|Y_0|$ = 8.7 µS, (b) RM2 $|Y_0|$ = 25.9 µS, (c) RM3 with $|Y_0|$ = 47.8 µS and (d) RM4 with $|Y_0|$ = 86.1 µS. The resonance frequency $f_r$ and the quality factor $Q$ at the magnetic working point extracted from the mBvD model are given in the top left corners of the subfigures.

## 2.5. Sensor Arrays

To demonstrate the high reproducibility of the resonators achieved with the presented production method, we examined two arrays of 10 (array ID1) and 14 parallel-connected sensor elements (array ID2). The sensors have in-plane dimensions of 640 µm × 90 µm (array ID1) and 430 µm × 100 µm (array ID2). The resonance frequencies of each sensor in the arrays were measured with a vibrometer. The standard deviation of the resonance frequency variation for the first three resonance modes RM1-3 of 24 sensors is approximately 0.13 %, which is 40-50 times smaller than previously investigated magnetoelectric sensors where the magnetic layer was deposited before releasing the resonator.[30,32] In our resonators, RM1 has the highest standard deviation of ≈ 0.16 %, followed by RM3 with 0.13 % and RM2 with 0.09 %. A histogram of the normalized resonance frequency deviation $\Delta f_r / f_{r,\text{mean}}$ is shown in **Figure 7**a. It comprises all resonance frequency data from RM1-3 of both arrays. A histogram of the bandwidth normalized resonance frequency deviation $\Delta f_{r,\text{BW}}$ ($\Delta f_{r,\text{BW}} = \Delta f_r / f_{\text{BW}} \approx \Delta f_r / f_r \cdot Q$)[29] is shown in Figure 7b. Approximately 70% of the resonators have $f_r$ within the $\Delta f_{r,\text{BW}} <$ 0.5, which is necessary to improve sensor's detection limits by noise averaging.[29]

The admittance characteristic of array ID1 with 10 parallel connected sensor elements is measured around RM3 at magnetic flux densities from -20 mT to 20 mT. Since the resonance frequencies are very similar, we can define an average resonance frequency of the array via an mBvD fit as well. The resulting normalized resonance frequency $f_r(B)/f_{r,\text{max}}$ is shown in Figure 5as a function of the applied magnetic flux density $B$. It follows the same w-shape as the individual sensor element analyzed in the previous sections with a similar minimum of $f_{r,\text{min}}/f_{r,\text{max}} \approx 99.75\,\%$. As expected, also the maximum relative magnetic sensitivity is similar with $S_{m,r} \approx 4.8\,\text{T}^{-1}$ at $B = -0.52$ mT (Figure 5c and Supporting Information).

The normalized admittance magnitude $|Y|/|Y_{\text{max}}|$ and relative electric sensitivity $S_{\text{el,r}}$ at this magnetic operating point are shown in Figure 7d. With a value of $S_{\text{el,r}} = 88.2$ mS at $f_r = 688$ kHz, the relative electrical sensitivity of the array is a factor of 4.5 times larger compared to the single sensor element (Section 2.4.2, RM3).

The total amplitude sensitivity at the magnetic working point reaches $S_{\text{am}} = 423\,\mu\text{S}\,\text{mT}^{-1}$, which is approximately 3.5 times larger than with the single sensor element and previously investigated mm-sized ΔE-effect sensors.[21,23,24,27] Overall, the results demonstrate the high reproducibility of the sensors due to excellent stress control with the shadow-mask deposition technology and its potential for sensor arrays.

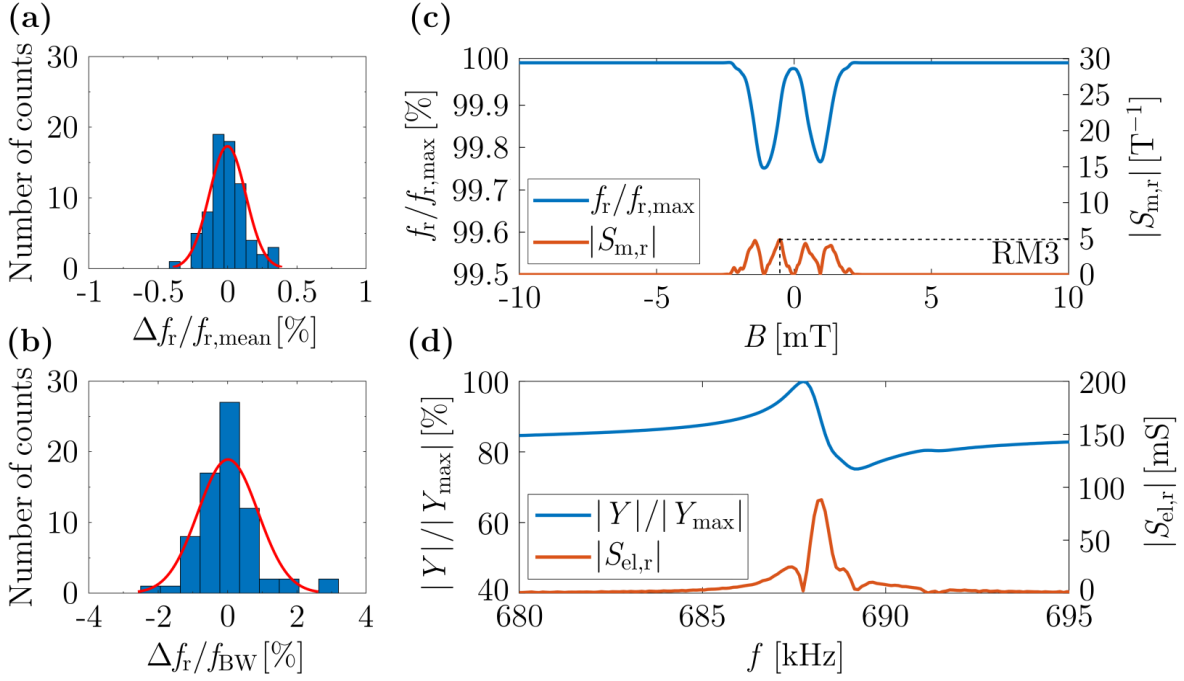

**Figure 7.** (a) Histogram of all normalized resonance frequency deviations $\Delta f_\mathrm{r}/f_\mathrm{r,mean}$, defined via the deviation $\Delta f_\mathrm{r}$ of the resonance frequency $f_\mathrm{r}$ from the mean resonance frequency $f_\mathrm{r,mean}$, and (b) histogram of the resonance frequency deviations, normalized to the averaged resonator bandwidth $f_\mathrm{BW}$ of the resonators for RM1-RM3 in array ID1 (10 resonators) and ID2 (14 resonators). (c) Normalized (mean) resonance frequency $f_\mathrm{r}/f_\mathrm{r,max}$ with $f_\mathrm{r,max} = 689$ kHz and relative magnetic sensitivity $S_\mathrm{m,r}$ as a function of the magnetic flux density $B$ applied along the long axis of the resonator for RM3 of array ID1, and (d) normalized admittance magnitude $|Y|/|Y_\mathrm{max}|$ with $Y_\mathrm{max} = 348$ μS and relative electrical sensitivity $S_\mathrm{el,r}$ of array ID1 at the working point ($B = -0.52$ mT) as functions of the frequency $f$.

## 3. Conclusion

We presented a shadow mask deposition technology combined with a free-free magnetoelectric microresonator design for miniaturized ΔE-effect sensors. The deposition of the magnetic layer through the shadow-mask avoids residual anisotropic stress from the nonmagnetic layers during microfabrication. The influence of the magnetic layer deposition on the anisotropic stress and magnetoelastic anisotropy was determined by combining a magneto-mechanical model and (magneto-)optical measurements. A small and homogeneous magnetoelastic anisotropy was achieved ($< 500$ Jm$^{-3}$). On the anchors, the demagnetizing field locally increases the effective anisotropy energy density and its inhomogeneity.

The first four resonance modes (RM) were analyzed with FEM simulations and vibrometer measurements. Frequency detuning via the ΔE effect, quality factors, electric sensitivities, and magnetic sensitivities are overall comparable with mm-sized ΔE-effect sensors,[21–24,26,27,29] despite the miniaturization. Owing to the miniaturized design, larger resonance frequencies between 125.1 kHz and 1.3 MHz are achieved, resulting in significantly higher resonator bandwidths from 0.2-3.4 kHz.

The asymmetric bending mode RM3 was identified as a particularly suitable resonance mode for device operation. It avoids the unfavorable magnetic layer properties on the resonator anchors, can be well excited electrically, and its large resonance frequency results in a higher resonator bandwidth than RM1. In RM3, a maximum amplitude sensitivity of $S_{am} = 121$ μS was measured with a bandwidth of 842 Hz, a relative electric sensitivity of $S_{el,r} = 19.5$ mS, and a relative magnetic sensitivity of $S_{m,r} = 6.2$ T$^{-1}$.

Arrays of parallel-connected magnetoelectric resonators demonstrated exceptional reproducibility with standard deviations < 0.2 % of resonance frequencies. This is 40-50 times smaller than previously investigated magnetoelectric sensors.[30,32] Using arrays of parallel connected sensor elements, the electric and total amplitude sensitivities were improved significantly by a factor of 4.5 and 3.5, respectively, compared to single sensor elements.

The results demonstrate promising progress in the fabrication technology for highly reproducible magnetoelastic structures and devices. Combined with the resonance modes identified, the microresonator design offers multiple benefits compared to traditional cantilevers and marks a notable step toward miniaturized ΔE-effect sensor arrays.

## 4. Experimental Section

*Device fabrication:* The magnetoelectric MEMS resonators were designed at Kiel University and fabricated at MEMSCAP Inc. by using a 5-mask level silicon-on-insulator (SOI) patterning and etching process. The magnetostrictive layers were deposited in a final step at Kiel University. The fabrication of the resonators starts from 150-mm-wide double-side polished (100)-oriented SOI wafer. The SOI wafer consists of 400-μm-thick substrate, 10-μm-thick polysilicon, and 1-μm-thick oxide layer. The poly-silicon layer is doped to serve as the bottom electrode. It is patterned and etched down to the oxide layer. An 0.2-μm-thick thermal oxide layer is grown and patterned to electrical isolate the doped poly-silicon layer from the following layers. An 0.5-μm-thick piezoelectric AlN layer is deposited by reactive sputtering and is then patterned and wet etched. After the deposition of the AlN layer, a stack of 20-nm-thick Cr and 1-µm-thick Al is deposited and patterned through a liftoff process. Polysilicon is patterned and etched by deep reactive ion etching (DRIE) down to the oxide layer. A polyimide coat is applied to the top surface of the patterned poly-silicon layer to keep the wafer together during the trench etching. After the reversal of the wafer, the bottom side of the substrate is patterned and etched to the bottom side oxide layer by reactive ion etching (RIE). The substrate layer is etched to the oxide layer by DRIE. The oxide layer in the area of the defined trench is removed by wet etching, and finally, the protective polyimide is removed by dry etching. Further details about the microfabrication process can be found elsewhere. [38]

*Magnetic layer deposition:* FeCoSiB with a thickness of 200 nm was deposited on the released resonators from their rear side with 10 nm Ta adhesion and capping layers. A shadow mask of the same size as the chip was placed directly on the rear side of the chip. The deposition was carried out at a working pressure of $3 \times 10^{-3}$ mbar with an Ar gas flow of 38 sccm and a power of 20 W. The deposition conditions were chosen to achieve the minimum stress arising from the deposition process. During the deposition, a magnetic field of 130 mT was applied by using two $Nd_2Fe_{14}B$ permanent magnets (Figure 1c) with dimensions of 22 mm × 8 mm × 3 mm to induce a uniaxial magnetic anisotropy perpendicular to the long axis of the resonators. After placing the chip and the mask, the magnets and the frame are covered by a top cover to avoid magnetic layer deposition on the sample holder and the magnets.

*Magnetic characterization:* Magnetic properties of the sensors were analyzed using magneto-optical Kerr effect (MOKE) microscopy.[34] To illustrate the domain configuration of the sample in the ground state (Figure 3a), the sample was demagnetized in a decaying sinusoidal magnetic field applied along its long axis (x-axis). The magneto-optical sensitivity axis was aligned along the short axis of the resonator (y-axis). The quasistatic magnetization curves in Figure 3c were recorded with the external magnetic field and the sensitivity axis oriented along the x-axis. The distribution of the differential magnetic susceptibility shown in Figure 3b was estimated from local MOKE magnetization curves like the ones shown in Figure 3c. The data are then compared with a macrospin magnetization model. The model considers magnetoelastic anisotropy energy, demagnetizing-field energy, uniaxial magnetization-induced anisotropy energy, and Zeeman energy.[26,39] Local magnetization curves are then calculated by minimizing the energy of the macrospin considering local values of the stress-induced anisotropy and the demagnetizing field. A saturation magnetization of $M_s = 1.5$ T was used for the simulation.[33]

*Vibrometer measurements:* Vibrometer measurements were performed using a Polytec MSA-500 Micro System Analyzer to determine resonance mode shapes and out-of-plane displacements. The sensor was electrically excited with a sinusoidal voltage with an amplitude of 100 mV via one top electrode at $B = 0$ mT. The top layer of the sensor was scanned with a focused laser with a 10-times objective lens using a grid including 57 data points in total. Data points were selected on the Al electrodes and the conduction lines.

*Definition of the sensitivities:* As a measure for the sensitivity, we use the amplitude sensitivity, as defined previously[27]

$$S_{am} := \left.\frac{\partial |Y|}{\partial B}\right|_{B=B_0, f=f_r} = S_{m,r} \cdot S_{el,r} \quad (1)$$

with the electric sensor admittance $Y$ relative magnetic sensitivity $S_{m,r}$ and the electric sensitivity $S_{el,r}$,

$$S_{m,r} = \left.\frac{1}{f_r}\frac{\partial f_r}{\partial B}\right|_{B=B_0}, \quad S_{el,r} = \left.\frac{\partial |Y|}{\partial f}\right|_{f=f_r} \cdot f_r \quad (2)$$

with the resonance frequency $f_r$, the magnetic bias flux density $B = B_0$ for an operating frequency $f = f_r$.

*Resonance frequency detuning:* To determine the resonance frequency as a function of the magnetic bias flux density *B* shown in Figure 5 and Figure **7**, we measured the admittance magnitude as a function of excitation frequency $f_{ex}$ for the magnetic flux density applied along the long axis of the sensor starting from -20 mT to 20 mT and back. Admittance measurements were carried out with an Agilent 4294A Precision Impedance Analyzer. The resonators were excited via one of the top electrodes with an excitation amplitude of 50 mV. We selected $u_{ex} = 50$ mV because resonator nonlinearities set in at large excitation voltage, which leads to a reduction in electrical sensitivity (see Supporting Information). The resonance frequencies $f_r$ and quality factors $Q$ are obtained by fitting a modified Butterworth-van-Dyke (mBvD) model to the admittance measurements.[40]

*Finite element method simulations:* All finite element simulations were performed in COMSOL Multiphysics® v. 6.0[41] with the material parameters and layer thicknesses provided in **Table 1**. The in-plane dimensions and anchor geometry are provided in Table S1 (Supporting Information, Sensor ID1). The model geometry comprises all layers except for the negligible thin oxide layer and Cr spacers in the electrodes. For all mechanical simulations, we solve the linear mechanical equations of motion with fixed boundary conditions for the displacement at the anchors.

The resonance modes (Section 2.3) were calculated with an eigenfrequency study using isotropic damping obtained from the measured quality factors.

For the residual stress analysis (Section 2.2), the equilibrium stress $\boldsymbol{\sigma} = \boldsymbol{\sigma}_0 + \boldsymbol{C}:(\boldsymbol{\varepsilon} - \boldsymbol{\varepsilon}_0)$ and the equilibrium strain $\boldsymbol{\varepsilon}$ were calculated by solving the mechanical equation of motion for static equilibrium conditions. For the simulations, we used $\boldsymbol{\sigma}_0$ as a fitting parameter and set the initial strain $\boldsymbol{\varepsilon}_0$ to zero. First, a model geometry without a magnetic layer was used to fit the out-of-plane displacement $u_z$ to the measurements of the released resonator without magnetic layer. Here, the initial stress was applied to the Si substrate. Then, in a second model, the magnetic layer was added to the model geometry, and initial stress $\boldsymbol{\sigma}_0$ is applied to the FeCoSiB layer. The resulting $u_z$ of the sample is obtained by summing up the displacement fields from both simulations, which is justified by the linearity of the underlying equations.

The local demagnetizing field $H_D$ was calculated by solving the magnetostatic equations considering the magnetization of the whole sample aligned along the x-axis. The magnitude of the magnetization is equal to the saturation magnetization $M_s$. For all simulations, we consider $M_s = 1.5$ T. [33] Demagnetizing field energy can then be expressed as $U_D = -0.5\mu_0 H_{D,x} M_s m_x$, where $\mu_0$ is the magnetic permeability of vacuum, $H_{D,x}$ is the local x-axis component of the demagnetizing field. Here, we only consider the demagnetizing field along the x-axis to approximate the flux closure owing to the formation of closure domains.

Table 1. Young's modulus $E_m$ in magnetic saturation, Poisson's ratio $\nu$, density $\rho$, and thickness of the layers used in FEM models.

| Material | $E_m$ [GPa] | $\nu$ | $\rho$ [kgm$^{-3}$] | Thickness |
|---|---|---|---|---|
| Poly-Si | 160 | 0.22 | 2300 | 10 µm |
| FeCoSiB | 150 | 0.3 | 7600 | 200 nm |
| AlN | 330 | 0.24 | 3300 | 0.5 µm |
| Al | 70 | 0.33 | 2700 | 1 µm |

**Acknowledgements**

The authors would like to thank Stefan Rehders for the design and fabrication of the custom deposition chamber and technical assistance, Lars Thormählen and Dirk Meyners for the discussions on the magnetic layer deposition, and Jörg Albers for the support with the vibrometer measurements. The research was funded by the German Research Foundation (Deutsche Forschungsgemeinschaft, DFG) through the Collaborative Research Centre CRC 1261 "Magnetoelectric Sensors: From Composite Materials to Biomagnetic Diagnostics", and the Carl Zeiss foundation via the project Memwerk.

**References**

[1]   V. Apicella, C. S. Clemente, D. Davino, D. Leone, C. Visone, Actuators 2019, 8, 45.

[2]   M. Díaz-Michelena, Sensors (Basel, Switzerland) 2009, 9, 2271.

[3]   C. Treutler, Sensors and Actuators A: Physical 2001, 91, 2.

[4]   I. R. McFadyen, E. E. Fullerton, M. J. Carey, MRS Bulletin 2006, 31, 379.

[5]   G. Lin, D. Makarov, O. G. Schmidt, Lab on a chip 2017, 17, 1884.

[6]   D. Murzin, D. J. Mapps, K. Levada, V. Belyaev, A. Omelyanchik, L. Panina, V. Rodionova, Sensors (Basel, Switzerland) 2020, 20.

[7]   E. Elzenheimer, C. Bald, E. Engelhardt, J. Hoffmann, P. Hayes, J. Arbustini, A. Bahr, E. Quandt, M. Höft, G. Schmidt, Sensors (Basel, Switzerland) 2022, 22.

[8]   Y. Wang, J. Li, D. Viehland, Materials Today 2014, 17, 269.

[9]   J. Reermann, P. Durdaut, S. Salzer, T. Demming, A. Piorra, E. Quandt, N. Frey, M. Höft, G. Schmidt, Measurement 2018, 116, 230.

[10]   R.-M. Friedrich, S. Zabel, A. Galka, N. Lukat, J.-M. Wagner, C. Kirchhof, E. Quandt, J. McCord, C. Selhuber-Unkel, M. Siniatchkin, F. Faupel, Sci Rep 2019, 9, 2086.

[11]   S. Zuo, J. Schmalz, M.-O. Ozden, M. Gerken, J. Su, F. Niekiel, F. Lofink, K. Nazarpour, H. Heidari, IEEE transactions on biomedical circuits and systems 2020, 14, 971.

[12]   Y. Huo, S. Sofronici, X. Wang, M. J. D'Agati, P. Finkel, K. Bussmann, T. Mion, M. Staruch, N. J. Jones, B. Wheeler, K. L. McLaughlin, M. G. Allen, R. H. Olsson, IEEE Sensors J. 2023, 23, 14025.

[13]   M. Bichurin, R. Petrov, O. Sokolov, V. Leontiev, V. Kuts, D. Kiselev, Y. Wang, Sensors (Basel, Switzerland) 2021, 21.

[14]   H. Li, Z. Zou, Y. Yang, P. Shi, X. Wu, J. Ou-Yang, X. Yang, Y. Zhang, B. Zhu, S. Chen, IEEE Transactions on Magnetics 2020, 56, 1.

[15]   H. J. Kim, S. Wang, C. Xu, D. Laughlin, J. Zhu, G. Piazza, in MEMS 2017: The 30th IEEE International Conference on Micro Electro Mechanical Systems Las Vegas, Nevada, USA January 22-26 2017, IEEE. Piscataway, NJ 2017, p. 109.

[16]   Y. Wang, D. Gray, D. Berry, J. Gao, M. Li, J. Li, D. Viehland, Advanced materials (Deerfield Beach, Fla.) 2011, 23, 4111.

[17]   V. Röbisch, E. Yarar, N. O. Urs, I. Teliban, R. Knöchel, J. McCord, E. Quandt, D. Meyners, Journal of Applied Physics 2015, 117.

[18]   E. W. Lee, Rep. Prog. Phys. 1955, 18, 184.

[19]   J. D. Livingston, phys. stat. sol. (a) 1982, 70, 591.

[20]   J. Reermann, S. Zabel, C. Kirchhof, E. Quandt, F. Faupel, G. Schmidt, IEEE Sensors J. 2016, 16, 4891.

[21]   S. Zabel, C. Kirchhof, E. Yarar, D. Meyners, E. Quandt, F. Faupel, Applied Physics Letters 2015, 107.

[22]   S. Zabel, J. Reermann, S. Fichtner, C. Kirchhof, E. Quandt, B. Wagner, G. Schmidt, F. Faupel, Applied Physics Letters 2016, 108.

[23]   P. Durdaut, J. Reermann, S. Zabel, C. Kirchhof, E. Quandt, F. Faupel, G. Schmidt, R. Knochel, M. Hoft, IEEE Trans. Instrum. Meas. 2017, 66, 2771.

[24]   B. Spetzler, C. Bald, P. Durdaut, J. Reermann, C. Kirchhof, A. Teplyuk, D. Meyners, E. Quandt, M. Höft, G. Schmidt, F. Faupel, Sci Rep 2021, 11, 5269.

[25]   T. Nan, Y. Hui, M. Rinaldi, N. X. Sun, Sci Rep 2013, 3, 1985.

[26]  B. Spetzler, C. Kirchhof, E. Quandt, J. McCord, F. Faupel, Phys. Rev. Appl. 2019, 12, 64036.

[27]  B. Spetzler, E. V. Golubeva, R.-M. Friedrich, S. Zabel, C. Kirchhof, D. Meyners, J. McCord, F. Faupel, Sensors (Basel, Switzerland) 2021, 21.

[28]  E. Spetzler, B. Spetzler, J. McCord, Adv Funct Materials 2023.

[29]  B. Spetzler, P. Wiegand, P. Durdaut, M. Höft, A. Bahr, R. Rieger, F. Faupel, Sensors (Basel, Switzerland) 2021, 21.

[30]  A. D. Matyushov, B. Spetzler, M. Zaeimbashi, J. Zhou, Z. Qian, E. V. Golubeva, C. Tu, Y. Guo, B. F. Chen, D. Wang, A. Will-Cole, H. Chen, M. Rinaldi, J. McCord, F. Faupel, N. X. Sun, Advanced materials technologies 2021, 6.

[31]  E. Yarar, S. Salzer, V. Hrkac, A. Piorra, M. Höft, R. Knöchel, L. Kienle, E. Quandt, Applied Physics Letters 2016, 109.

[32]  J. Su, F. Niekiel, S. Fichtner, L. Thormaehlen, C. Kirchhof, D. Meyners, E. Quandt, B. Wagner, F. Lofink, Applied Physics Letters 2020, 117.

[33]  A. Ludwig, E. Quandt, IEEE Transactions on Magnetics 2002, 38, 2829.

[34]  J. McCord, J. Phys. D: Appl. Phys. 2015, 48, 333001.

[35]  R. C. O'Handley, Modern magnetic materials: Principles and applications, Wiley, New York, Weinheim 2000.

[36]  P. Durdaut, E. Rubiola, J.-M. Friedt, C. Muller, B. Spetzler, C. Kirchhof, D. Meyners, E. Quandt, F. Faupel, J. McCord, R. Knochel, M. Hoft, J. Microelectromech. Syst. 2020, 29, 1347.

[37]  B. Spetzler, C. Kirchhof, J. Reermann, P. Durdaut, M. Höft, G. Schmidt, E. Quandt, F. Faupel, Applied Physics Letters 2019, 114.

[38]  A. Cowen, G. Hames, K. Glukh, B. Hardy, PiezoMUMPs(TM) Design Handbook, MEMSCAP Inc. 2014.

[39]  E. C. Stoner, E. P. Wohlfarth, IEEE Transactions on Magnetics 1991, 27, 3475.

[40]   K. S. van Dyke, Proc. IRE 1928, 16, 742.

[41]   COMSOL Multiphysics (TM) v. 6.0, COMSOL AB, Stockholm, Schweden 2018.

# Supporting Information

**Miniaturized Double-Wing Delta-E Effect Sensors**

*Fatih Ilgaz, Elizaveta Spetzler, Patrick Wiegand, Robert Rieger, Jeffrey McCord, Franz Faupel, Benjamin Spetzler\**

\*Correspondence: benjamin.spetzler@tu-ilmenau.de

## S1. Equilibrium stress in the magnetic layer after deposition

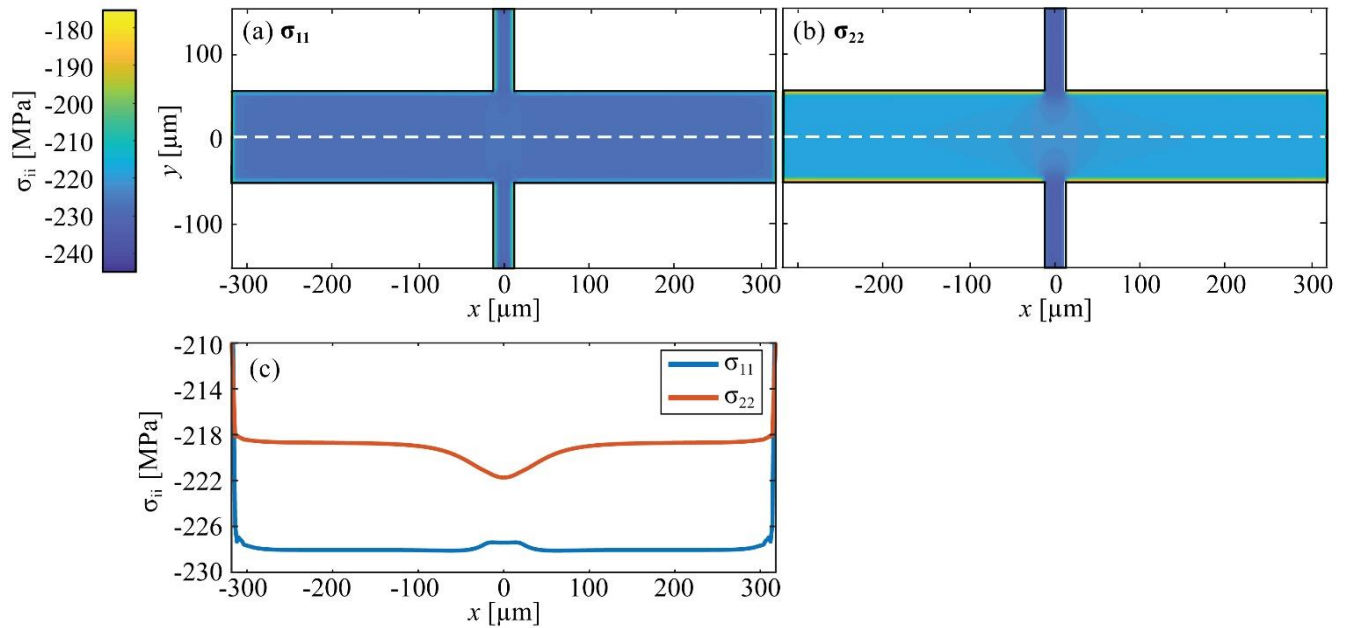

**Figure S1.** Spatial distribution of the equilibrium stress components in the center plane of the magnetic layer simulated with the FEM model. Initial stress $\sigma_{11} = -245$ MPa and $\sigma_{22} = -235$ MPa was applied to the whole volume of the magnetic layer. (a)-(b) Spatial distribution of the equilibrium stress components $\sigma_{11}$ and $\sigma_{22}$. (c) Distribution of the equilibrium $\sigma_{11}$ and $\sigma_{22}$ along the cut line marked with the white dashed line in (a) and (b).

## S2. Spatial distribution of the magnetic anisotropy

The simulated magnetic anisotropy from the demagnetizing field is shown in Fig. S1a,b, and the stress anisotropy $\sigma_{11} - \sigma_{22}$ after relaxation in the center of the magnetic layer in Fig. S1c,d.

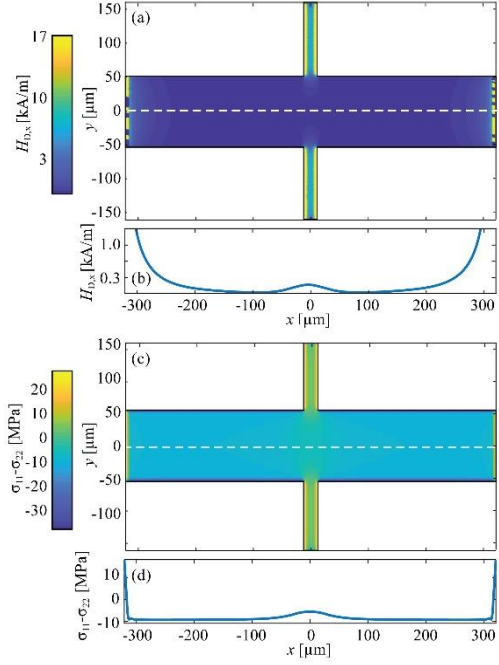

**Figure S2.** Simulated with FEM models spatial distribution of magnetic properties. (a)-(b) Distribution of the x-axis component of the demagnetizing field $H_{D,x}$. (c)-(d) Anisotropy $\sigma_{11} - \sigma_{22}$ of the in-plane stress in the magnetic layer after deposition in equilibrium. (b) and (c) show a cut line through the plots in (a) and (c), respectively.

## S3. Magnetic Properties of Different Sensor Geometries

In this study, we investigated twelve single sensors with varying geometries, i.e., in-plane dimensions and anchor width, and two sensor arrays with identical ten and fourteen parallel-connected sensors from the same chip while focusing on one sensor (Sensor ID1) in detail. Below, magnetic domain images in a demagnetized state (Figure S3) and magnetization curves (Figure S4) of the single sensors (ID2-12) can be found. Detailed magnetic properties of ID1 are given in the main text.

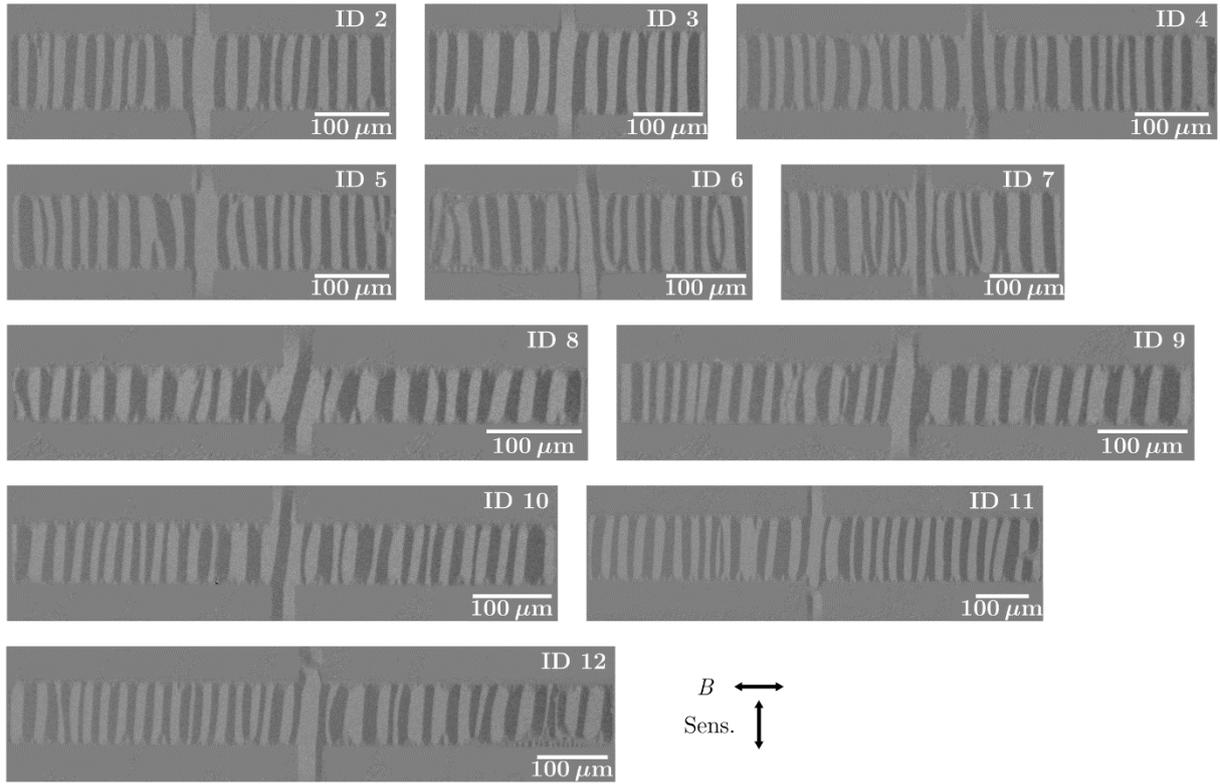

**Figure S3.** Magnetic domain images of the investigated sensors (ID2-12) after demagnetizing the sensor along their long axes. Magneto-optical sensitivity is aligned perpendicular to the demagnetizing field.

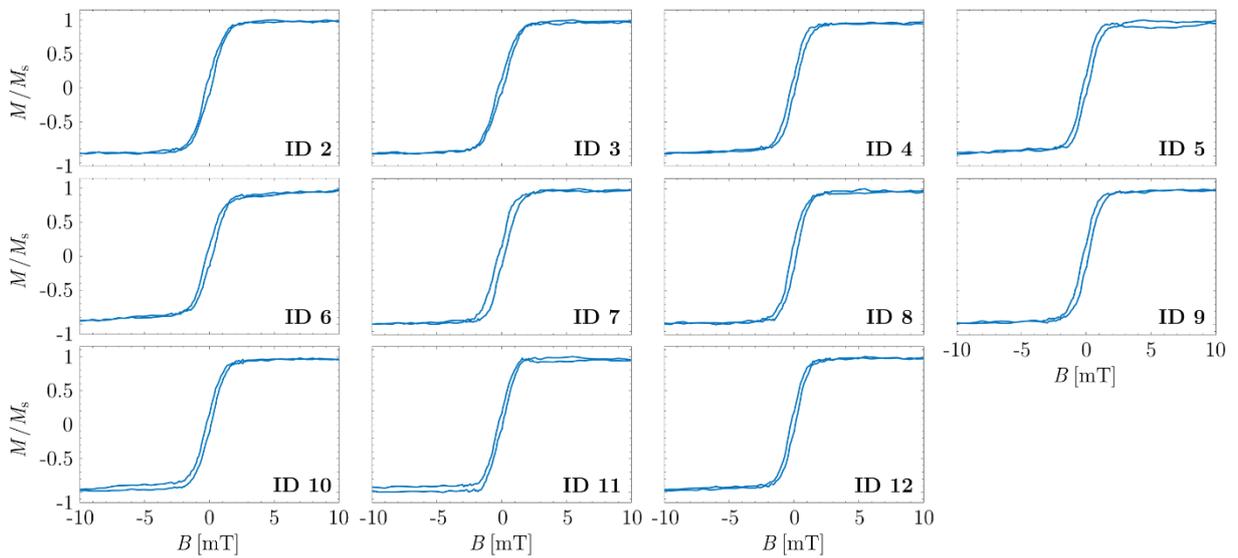

**Figure S4.** Magnetization curves of the investigated sensors (ID2-12) measured along their long axes.

## S4. Nonlinearity

We measured the admittance magnitude of the sensor ID1 at its magnetic working point and different excitation voltage amplitudes and determined the electrical sensitivity. Figure S5a shows that the resonance peak in the admittance curve shifts toward smaller frequencies with increasing excitation voltage amplitude $u_{ex}$, with reduced electrical sensitivity, which becomes significantly visible above 100 mV. To understand the origin of this nonlinearity, admittance magnitude measurements were also done at the magnetic saturation (Figure S5b). In contrast to measurements at the magnetic working point, admittance magnitude and electrical sensitivities are independent of the excitation voltage amplitude. It shows that nonlinearity results from a magnetostrictive origin. An excitation voltage amplitude of 50 mV was used for all measurements in this study since the sensors are still in a linear regime with sufficient signal amplitude.

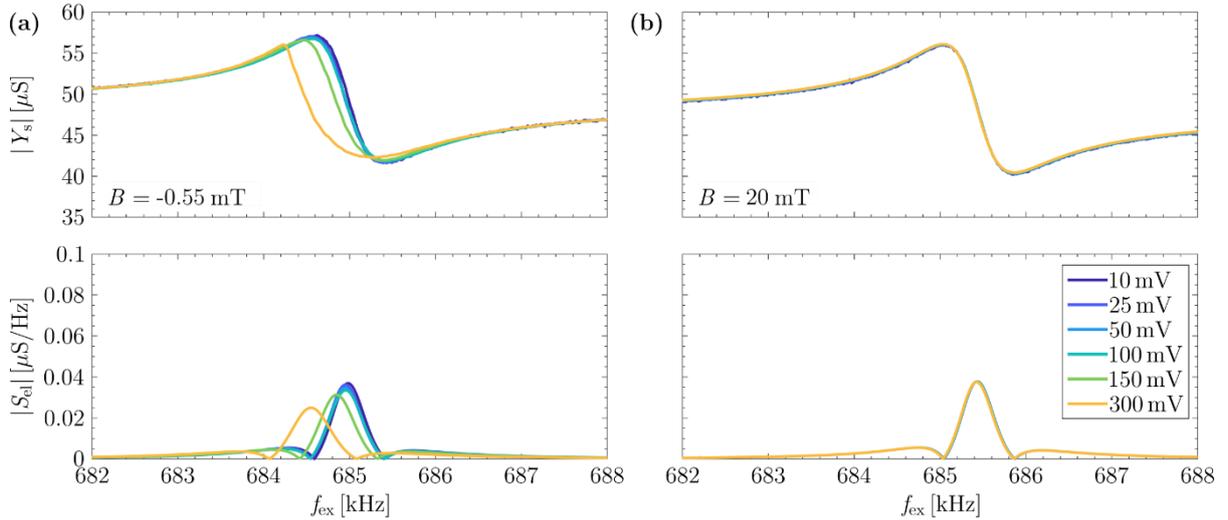

**Figure S5.** (a) Admittance magnitude $|Y_s|$ and magnitude of electrical sensitivity $|S_{el}|$ of sensor ID1 for RM3 at different excitation voltage amplitudes and magnetic bias field of $B = -0.55$ mT and (b) $B = 20$ mT.

## S5. Stress Components in RM1-RM4

The stress magnitudes of the four resonance mode shapes are shown in Figure S6. In RM3, a significant contribution of the shear stress component $\sigma_{12}$ is apparent.

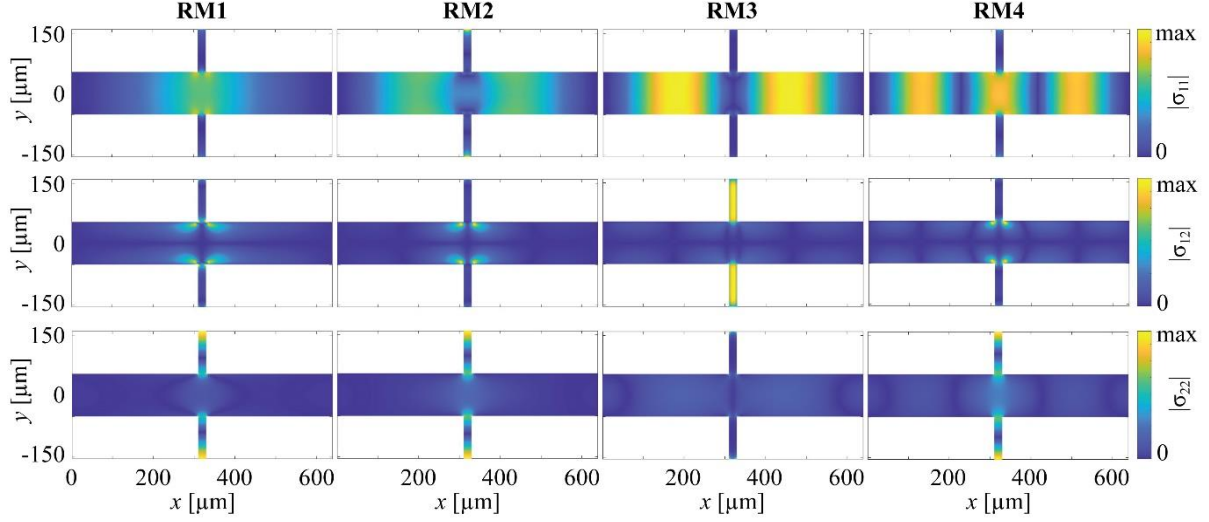

**Figure S6.** Magnitudes of the three stress components $\sigma_{11}$, $\sigma_{12}$, $\sigma_{22}$ in the four resonance modes RM1-4 in the center of the magnetic layer.

## S6. Sensor Characteristics

Resonance frequencies, sensitivities, and quality factors for RM1-4 of the single sensors, except for RM1 of Sensor ID3 and ID11 (due to noise) and RM1,2 of ID4 (due to a broken sensor), are determined. The geometries and the resonance frequencies $f_r$ of the investigated single sensors are given in Table S1. Relative magnetic, relative electrical, and total amplitude sensitivities are shown in Table S2. Quality factors at zero bias field can be found in Table S3. Table entries that could not be determined due to noise (RM1 of ID3 and ID11) and a broken resonator (RM1 and 2 of ID4) marked with "X".

**Table S1.** Parameters of the investigated sensors, including in-plane dimensions, anchor widths, and measured resonance frequencies $f_r$ of the first four resonance modes RM1-4 at zero magnetic bias field.

| Sensor ID | In-plane dimensions [μm ×μm] | Anchor width [μm] | $f_r$ [kHz] | | | |
|---|---|---|---|---|---|---|
| | | | RM1 | RM2 | RM3 | RM4 |
| 1 | 640 × 105 | 25 | 125.7 | 365.9 | 685.2 | 1301.8 |
| 2 | 510 × 105 | 25 | 171 | 479.7 | 1072.1 | 1985 |
| 3 | 400 × 125 | 25 | X | 676.1 | 1734 | 2692 |
| 4 | 640 × 105 | 30 | X | X | 690.8 | 1298.6 |
| 5 | 510 × 105 | 30 | 176.9 | 490.8 | 1080.5 | 1985.4 |
| 6 | 400 × 100 | 30 | 235.1 | 699.5 | 1747.3 | 2792.7 |
| 7 | 400 × 125 | 30 | 210 | 681.6 | 1741 | 2643.4 |
| 8 | 600 × 60 | 30 | 157.1 | 488.9 | 798.5 | 1448.9 |
| 9 | 640 × 70 | 30 | 137.9 | 437.2 | 700.2 | 1295.3 |
| 10 | 690 × 80 | 30 | 114.4 | 340.3 | 598.3 | 1106.7 |
| 11 | 850 × 125 | 30 | X | 218.8 | 390.8 | 745.3 |

| 12 | 850 × 90 | 35 | 79.8 | 276 | 398.4 | 757.8 |

**Table S2.** Relative magnetic $S_{m,r}$, relative electrical $S_{el,r}$, and total amplitude sensitivities $S_{am}$ of the investigated sensors at their magnetic working points for RM1-RM4.

| Sensor ID | $S_{m,r}$ [T$^{-1}$] | | | | $S_{el,r}$ [mS] | | | | $S_{am}$ [µS mT$^{-1}$] | | | |
|---|---|---|---|---|---|---|---|---|---|---|---|---|
| | RM1 | RM2 | RM3 | RM4 | RM1 | RM2 | RM3 | RM4 | RM1 | RM2 | RM3 | RM4 |
| 1 | 8.9 | 2.7 | 6.2 | 4.8 | 0.8 | 4.0 | 19.5 | 2.8 | 7 | 11 | 121 | 14 |
| 2 | 7.4 | 3.0 | 6.9 | 4.3 | 0.4 | 10.3 | 12.5 | 2.0 | 3 | 31 | 86 | 9 |
| 3 | X | 6.4 | 5.2 | 2.4 | X | 4.7 | 4.9 | 1.2 | X | 30 | 25 | 3 |
| 4 | X | X | 5.6 | 4.3 | X | X | 20.8 | 2.2 | X | X | 116 | 9 |
| 5 | 7.6 | 2.9 | 5.5 | 4.9 | 0.6 | 9.0 | 14.7 | 2.1 | 5 | 25 | 81 | 10 |
| 6 | 3.7 | 2.6 | 3.4 | 1.1 | 0.2 | 8.9 | 6.5 | 3.4 | 1 | 23 | 22 | 4 |
| 7 | 4.6 | 3.6 | 3.8 | 1.5 | 0.0 | 7.6 | 5.8 | 6.3 | 0 | 27 | 22 | 9 |
| 8 | 4.9 | 1.7 | 4.1 | 4.1 | 0.9 | 2.4 | 11.9 | 2.0 | 4 | 4 | 49 | 8 |
| 9 | 7.5 | 2.0 | 6.4 | 4.2 | 0.7 | 3.8 | 14.9 | 4.5 | 5 | 8 | 95 | 19 |
| 10 | 5.8 | 2.3 | 6.2 | 5.0 | 1.1 | 3.9 | 13.7 | 4.2 | 6 | 9 | 85 | 21 |
| 11 | X | 3.1 | 8.2 | 6.1 | X | 2.5 | 12.7 | 4.2 | X | 8 | 104 | 26 |
| 12 | 9.9 | 3.1 | 8.2 | 5.5 | 0.7 | 2.6 | 12.6 | 2.8 | 7 | 8 | 103 | 15 |

**Table S3.** Quality factors of the investigated sensors at zero magnetic bias field for RM1-RM4.

| Q | Sensor ID | | | | | | | | | | | |
|---|---|---|---|---|---|---|---|---|---|---|---|---|
| | 1 | 2 | 3 | 4 | 5 | 6 | 7 | 8 | 9 | 10 | 11 | 12 |
| RM1 | 609 | 525 | X | X | 608 | 640 | 422 | 690 | 551 | 621 | X | 490 |
| RM2 | 687 | 828 | 572 | X | 732 | 681 | 604 | 683 | 744 | 670 | 459 | 594 |
| RM3 | 814 | 579 | 319 | 764 | 596 | 355 | 264 | 945 | 920 | 838 | 754 | 789 |
| RM4 | 394 | 264 | 342 | 363 | 265 | 515 | 517 | 512 | 689 | 624 | 508 | 605 |